\documentclass[journal]{IEEEtran}

\usepackage{graphicx}
\usepackage{subfigure}
\usepackage{cite}
\usepackage{multirow}
\usepackage{slashbox}
\usepackage{diagbox}
\usepackage{amssymb}
\usepackage{array}
\usepackage{slashbox}
\usepackage{gensymb}
\usepackage{bm}

\ifCLASSINFOpdf
\else
\fi
\begin{document}
%
\title{A Review on Serious Games for Phobia}
%
%
%

\author{Sha~Li, Peichen~Yang, Rongyang~Li, Fadi~Farha, Jianguo~Ding, Per~Backlund,\\and~Huansheng~Ning,~\IEEEmembership{Senior Member,~IEEE}
}

\maketitle

\begin{abstract}

Phobia is a widespread mental illness, and severe phobias can seriously impact patients’ daily lives. One-session Exposure Treatment (OST) has been used to treat phobias in the early days,but it has many disadvantages. As a new way to treat a phobia, virtual reality exposure therapy(VRET) based on serious games is introduced. There have been much researches in the field of serious games for phobia therapy (SGPT), so this paper presents a detailed review of SGPT from three perspectives. First, SGPT in different stages has different forms  with the update and iteration of technology. Therefore, we reviewed the development history of SGPT from the perspective of equipment. Secondly, there is no unified classification framework for a large number of SGPT. So we classified and combed SGPT according to different types of phobias. Finally, most articles on SGPT have studied the therapeutic effects of serious games from a medical perspective, and few have studied serious games from a technical perspective. Therefore, we conducted in-depth research on SGPT from a technical perspective in order to provide technical guidance for the development of SGPT. Accordingly, the challenges facing the existing technology has been explored and listed.
\end{abstract}


%
\IEEEpeerreviewmaketitle

\section{Introduction}
%
%
%
%
\IEEEPARstart{P}{hobia} is defined as an intense and persistent fear when patients face certain specific things or scenes \cite{RE-1}. The clinical physiological symptoms of fear include fainting, heart rate acceleration, cognitive bias, a drop in blood pressure, and attention bias \cite{RE-2}. The fainting, worrying, and avoidance behavior associated with phobias have a severe impact on a patient’s life and work \cite{RE-3}. The traditional treatment of phobias is one-session exposure therapy (OST), in which the patients systematically and repeatedly are exposed to specific things or situations for three hours \cite{RE-4,RE-5}. However, OST has also many limitations, such as the inability to control the level of fear in exposure scenarios, the patients are unwilling to face the real scenarios that scares them, and the inability to recreate a certain scenarios for exposure in reality \cite{RE-6}. Today, with the increasing popularity of virtual reality (VR) technology and the rapid evolution of computer technology, virtual reality exposure therapy (VRET) based on serious games is emerging as a more effective treatment \cite{RE-7}. As illustrated by Garcia-Palacios A et. al.\cite{RE-8}, VRET has a better effect in treating the fear of spiders compared with traditional OST. This suggests that VRET has great application prospect in the treatment of phobias.
\footnote{Sha Li, Peichen Yang, RongYang Li, Fadi Farha, Huansheng Ning is with the School of
	 Computer and Communication Engineering, University of Science and Technology Beijing, 10083, Beijing, China, e-mail:ninghuansheng@ustb.edu.cn.} 
\footnote{Jianguo Ding is with the School of Information, University of Skovde, Skovde, Sweden, e-mail:jianguo.ding@his.se.} 
\footnote{Per Backlund is with the School of Information, University of Skovde, Skovde, Sweden,
	 e-mail:per.backlund@his.se.}

Serious games has emerged as a new trend in phobia therapy, and many related studies were being conducted. In the beginning, serious games based on VR-BOX, such as Live Beyond Fear \cite{RE-9}. It have been used to treat acrophobia. With the maturity of VR technology, serious games based on Head-Mounted Displays (HMD) for phobia treatment are becoming mainstream, which exhibits good performance. To further enhance a sense of reality and immersion, serious games based on Cave Automatic Virtual Environment (CAVE) are also gaining ground among researchers \cite{RE-10}. Although the system is less widely used in SGPT, researchers have demonstrated its superiority \cite{RE-11}. The study of SGPT has been going on for a long time. But there is still a lack of a systematic overview of its development. Therefore, we have summarized the existing SGPT. The goal is to help researchers quickly understand its development in a short period of time.

After long-term of studies, researchers have proposed many serious games for different types of phobias. However, This is not conducive for medical staff or patients to choose suitable serious games to treat the disease. Given these problems, we refer to the criteria for the classification of phobias established by DSM-5 and analyze the characteristics of different types of phobias \cite{RE-12}. On this basis, we present a category of SGPT to offer reliable guidance for researchers. Besides, we hope these researchers will be able to design the most effective serious games based on the characteristics of different types of phobias to achieve an optimum match between game development and therapeutic purpose.

There has been a lot of research showing the effectiveness of serious games in phobias treatment. However, there are no literature reviews of the technologies used in SGPT. To address these issues, a technical analysis model is proposed in this paper. It analyzes the main technologies used in SGPT from a game development perspective. Thus, game development process is divided into three stages: design, development, and evaluation. The optimal therapeutic effect of SGPT is achieved by using different key technologies at each stage. 

In this paper, we first analyzed and described the history of SGPT, and categorized existing serious games based on virtual reality display device and clinical symptoms of phobias. Then, we concluded the major technologies through the combination of game development and SGPT technology. The rest of the paper is structured as follows: Section II intorduces the development history of SGPT, including serious games based on VR-BOX, serious games based on HMD, and serious games based on CAVE; Section III introduces the categories of SGPT based on characteristics of phobias, including serious games for phobias of thing, serious games for phobias of scene, and serious games for phobias of physiological phenomenon; Section IV proposes the technical analysis model; Section V discusses the challenges of SGPT and gives some advice; Finally, Section VI summarizes the whole passage and gives some relevant conclusions.

\section{The development of SGPT}
Virtual reality exposure therapy based on serious games is a perfect combination of serious games and virtual reality technology. Serious games originated in the education industry, which allows students to complete learning tasks in addition to entertainment in order to achieve the purpose of teaching and fun. With its growth, it is now used in a wide range of medical, engineering, scientific, political, and military industries. While traditional video games focus on entertainment and provide a fun and rich experience for the user, serious games are more "task" oriented. Virtual reality (VR) is a computer technology that allows users to interact with a computer-generated virtual world through lifelike images, sounds, I/O devices and other motion-sensing devices, giving users an immersive feeling. Virtual reality exposure therapy is based on virtual reality technology, and uses serious games as an application platform to achieve a specific "task"----phobia treatment.

\begin{figure*}[t!]
	\centering
	\includegraphics[width=15cm,height=7cm]{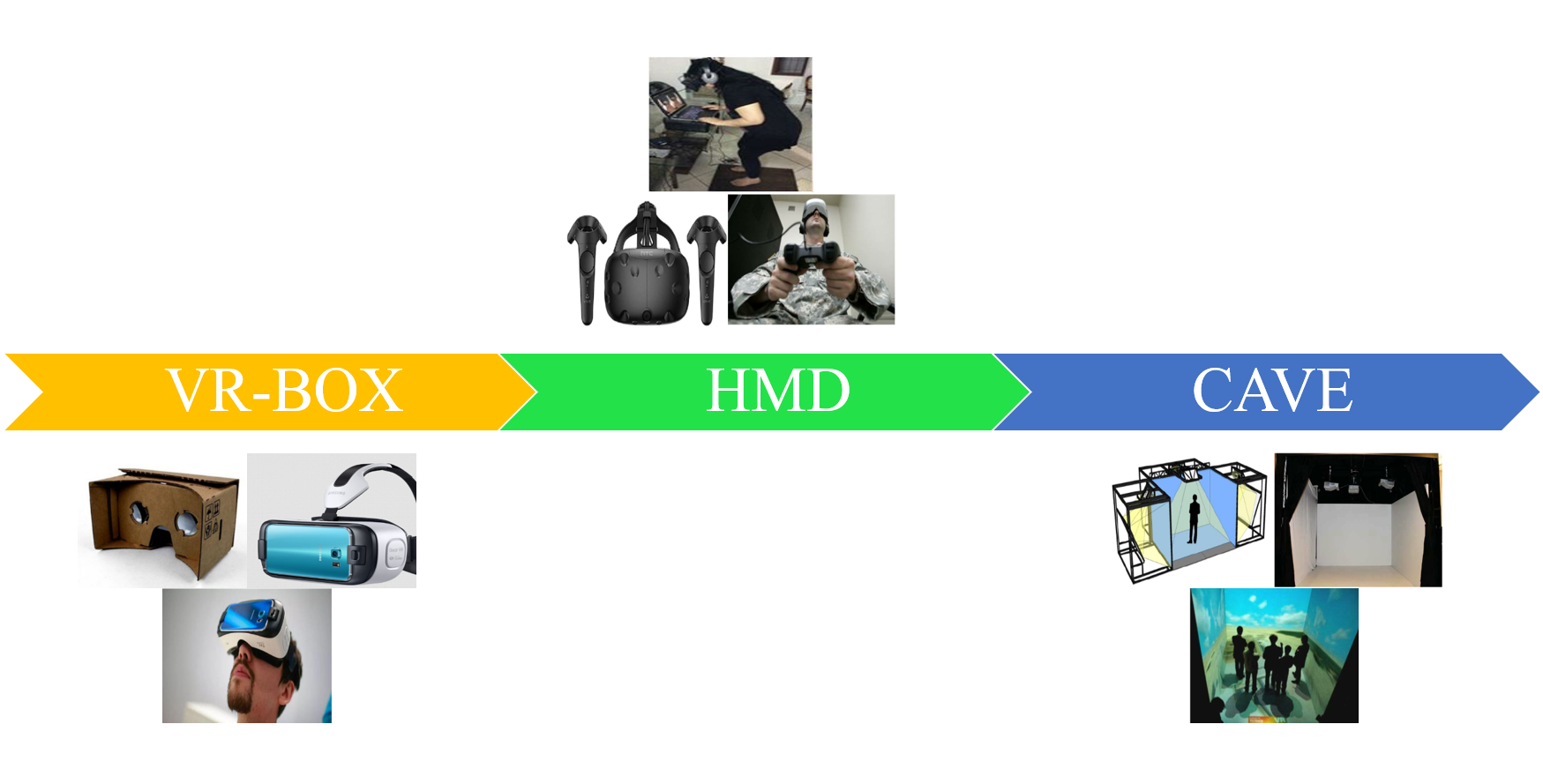}\\
	\caption{The development of SGPT }
	\label{Figure-1}
\end{figure*}

VRET based on serious games has shown great potential in phobia treatment. Therefore, understanding its development will help researchers gain a clearer cognizance of SGPT. As shown in Fig. \ref{Figure-1}, we reviewed SGPT development from the perspective of the VR Display Device. SGPT is divided into three stages: serious games based on VR-BOX, serious games based on HMD, and serious games based on CAVE. In addition to, we summarized related devices and features of these serious games for SGPT in Table \ref{Tab-1}.

\begin{table*}\normalsize
\newcommand{\tabincell}[2]{\begin{tabular}{@{}#1@{}}#2\end{tabular}}
\renewcommand\arraystretch{2}
	\centering
	\caption{Comparison among different kinds of games for SGPT}
	\label{Tab-1}
	\begin{tabular}{|m{3cm}<{\centering}|m{6cm}<{\centering}|m{8cm}<{\centering}|}\hline
		{Game Types}&{Related Devices}&{Features}\\\hline
		\tabincell{c}{Games based on\\VR-BOX}&\tabincell{c}{Google Cardboard,\\UPG.VR, etc.}&\tabincell{l}{Advantages: Low cost and easily implement.\\Disadvantages: Limitation of reality and immersion.}\\\hline
		\tabincell{c}{Games based on\\HMD}&\tabincell{c}{5DT HMD800, Oculus Rift,\\HTC Vive, etc.}&\tabincell{l}{Advantages: Immersion, Availability, and Widespread.\\Disadvantages: Bulky and Dizziness.}\\\hline
		\tabincell{c}{Games based on\\CAVE}&\tabincell{c}{NVIDIA-3D vision Pro glasses,\\Crystal Eyes LCD Shutter glasses, etc.}&\tabincell{l}{Advantages: Advanced, Flexible, Share, and Portable.\\Disadvantages: High cost and Immaturity.}\\\hline
			
	\end{tabular}
\end{table*}

\subsection{Serious games based on VR-BOX for phobia therapy}
In the early stage of virtual reality system application. Due to the immaturity of computer technology, virtual reality technology, and hardware equipment. Researchers adopted VR-Box as VR display device in the stage. As the name suggests, this type of VR display device is an empty box with a VR lens, which needs the help of a mobile phone to achieve a 3D effect and finally realizes the virtual reality in a similar sense.

Live Beyond Fear \cite{RE-9} uses Google Cardboard as a virtual display device. Although it's just a simple VR-box, the device can be combined with any smartphone equipped with KitKat to achieve virtual reality. Viewed from a first-person perspective. The game has three different levels and each representing a different altitude environment. Participants need to complete a specific task in each level to move to the next level. As the plot progresses, the level of difficulty and excitement of the task increases gradually. Room VR \cite{RE-43} is also a serious game that uses Google Cardboard as a virtual reality device to treat darkness phobia. The game is still played in the first person, with the patient moving freely in a dark, closed room, and it's innovative in that it emphasizes the importance of light and music. The background color of the game is bluish purple, because according to psychology, purple can stimulate participants' curiosity and enhance their desire to play the game. Background music is played throughout the game to enhance the immersion of the entire game. A cross line in the game channel locates the position of the user's line of sight, and the light will illuminate the position the user sees as the line of sight moves.


In general, in the early stage of virtual reality technology is not mature, VR serious games based on VR-BOX can be regarded as the best choice. Besides, it is cheap and easy to implement. However, it cannot achieve an effective sense of reality VR effect due to the limitation of technology or unable to give an effective feasibility analysis due to the lack of a large number of experimental verification.

\subsection{Serious games based on HMD for phobia therapy}
Head-Mounted Displays (HMD) is a visual display system based on a single user. The system is connected to a computer. It uses the way the brain processes images seen by the eyes to show two specially processed images from different angles to the patient's two eyes for a 3D effect. In addition, the system also contains sensors, somatosensory devices, and tracking devices, which can track the patient's motion state and physiological response in real-time. Using these devices, the interaction between the virtual scene and the patient is great realized. The patient is completely separated from the real world \cite{RE-20}.

Jarrell et al. \cite{RE-21} proposed a game has been used to treat Post-Traumatic Stress Disorder (PTSD). It uses 5DT HMD800 as a VR display device, which has an 800*600 resolution screen. The patient was asked to sit in a chair and walk through the virtual world using a USB handle device. During walking, the InterSense InertiaCube2 tracker tracks the patient's head and body status. Cristina Botella et al. \cite{RE-22} proposed a serious game for the treatment of flight phobia that uses I-glasses as a VR display device. The device has a 640*480 resolution display screen. The device has a light body and a comfortable ergonomic design, which gives patients the best experience. Patients were also asked to sit in a chair and interact with the virtual environment through the mouse, whose interactive actions were captured through the InterSense 3D. Besides, Jarrell Pair et al. \cite{RE-23} proposed a serious game for the treatment of spider phobia, and Jessica et al. \cite{RE-24} proposed a serious game used to treat driving phobia. They all use I-glasses as VR display device.

Using the mouse, handles, and other wired devices achieve interaction with the virtual environment. There is also a class of HMDs that use a variety of somatosensory devices and sensors to interact with virtual environments, which gives patients greater activity space. Sherazade Shunnaq et al. \cite{RE-25} proposed a serious game model used to treat both acrophobia and spider phobia. Its VR display device is Oculus Rift with a resolution of 1280*800. Patients interact with the virtual environment through a somatosensory controller connected to the HMD. In addition to the Oculus Rift, the HTC vive is also one of the most popular HMDs. It has a 2K screen resolution and uses two wireless controllers to interact with the virtual environment. For example, Jessica S. Ortiz at el. \cite{RE-26} proposed a serious game that can be used to treat variously phobias, including spider phobia, acrophobia, and claustrophobia. Miroslav Musalek et al. \cite{RE-27} proposed a serious game to treat spider phobia. Their VR display devices all use HTC vive.

To sum up, the use of HMD in serious games based on VR technology has greatly promoted the development of VRET. Not only does it has a better sense of reality and immersion, but also a variety of new forms and categories. Besides, the system can better motivate patients and improve the acceptance rate of phobia treatment. Finally, a large number of research results have proved the effectiveness of the system.

Although VR serious games based on HMD have many advantages in treating phobias, they also have some limitations. HMD needs to be connected to Personal Computer(PC), causing limitation in patients' range of movement. In addition, most HMD are bulky and have a strong sense of immersion, so long-term play will cause a great burden on the body and lead to dizziness and nausea.
\subsection{Serious games based on CAVE for phobia therapy}
Serious games based on HMD will greatly limit the range of patients' activity. Hence, researchers proposed a new method, namely Computer Automatic Virtual Environment (CAVE). It is a virtual reality system based on multi-user projection. The projector is controlled by a PC with a professional graphics card, and the images from the computer are projected into a room, and finally, the room is built into a virtual reality scene. Patients interact with virtual reality scene by wearing a specific device. Because the equipment do not need to be connected to the PC, so the user can do free activities in the virtual scene \cite{RE-28}.

Joao p. Costa et al. \cite{RE-11} proposed a serious game for the treatment of acrophobia. The game of VR display device adopts the NVIDIA 3D Vision Pro glasses. The real-time tracking of the user's location is achieved through the infrared tracking tag on the user's hat. Meanwhile, the game also uses 3D rendering technology to enhance the immersion of the entire game environment. The game set up five different height environment, when the patient in the current high sense of fear and relaxation reached a certain level, will unlock the next height scene. Similarly, the serious game proposed by Daniel Gromer et al. \cite{RE-29} for the treatment of acrophobia is also based on CAVE. In the game, the developers used six high-resolution projectors to project images onto four walls and floors. Participants interact with the virtual environment by wearing interference-filtering glasses. Participants were asked to climb watchtowers to complete tasks that were fraught with danger and challenge.

In short, CAVE offers a higher sense of reality with the help of a super-high resolution projector than a VR serious game based on HMD. Besides, CAVE supports multi-person sharing of the virtual scene, and users are free to explore and enhance their immersion with accessories such as wireless controllers, joystick, or other virtual objects. Besides, CAVE supports multi-person sharing of the virtual scene. However, CAVE is expensive to treat. As a system, CAVE uses truly cutting-edge technology, which makes it complex to implement. In addition, the application of this system in the treatment of phobias is still in infancy, and there is a lack of effective research results.

\section{Serious games classification based on different types of phobia}
In this chapter, we referred to the Diagnostic and Statistical Manual of Mental Disorders-5th Edition (DSM-5) criteria for phobias. We proposed a classification model based on the clinical symptoms of different phobias \cite{RE-30}. The purpose of our classification is to give a clear guide about developing targeted serious games.That can help improve the effectiveness and efficiency of treatment by including the existing SGPT into the system to be concisely and efficiently in the future. The classification in \cite{RE-26} is too concise to fully include the existing SGPT. On the other hand, the classification in \cite{RE-19} is too detailed, which increases the complexity of system development. Therefore, compared to the classification in literature \cite{RE-26} and \cite{RE-19}, our classification is more neutral and reduces the complexity of game development while ensuring that all phobias are included as far as possible. SGPT is divided into three categories. TABLE \ref{Tab-2} depicts the SGPT classification of clinical symptoms based on different phobias.

\begin{table*}\normalsize
\newcommand{\tabincell}[2]{\begin{tabular}{@{}#1@{}}#2\end{tabular}}
\renewcommand\arraystretch{1.8}	
	\centering
	\caption{Serious games for different phobias types using different devices}
	\label{Tab-2}
	\begin{tabular}{|c|m{3cm}<{\centering}|m{3cm}<{\centering}|m{3cm}<{\centering}|m{3cm}<{\centering}|}\hline
		Phobia Types&Therapeutic Phobias&Related Literature&Games&Devices Types\\\hline
		
		\multirow{6}{*}{Phobias of object}
		&\multirow{3}{*}{Fear of Spider}&Philip Lindner et al.\cite{RE-31}&Itsy&VR-BOX\\\cline{3-5}
		&\multirow{3}{*}&ALBERT S. CARLIN et al.\cite{RE-32}&Spider World&HMD\\\cline{3-5}
		&\multirow{3}{*}&David Bel Lang et al.\cite{RE-33}&Lo-Fi-Prototype&CAVE\\\cline{2-5}
		&Cockroach Phobia&Juan, M.C. et al. \cite{RE-34}&ARcockroach&HMD\\\cline{2-5}
		&Blood Phobia&João Petersen et al. \cite{RE-35}&First Insights&HMD\\\cline{2-5}
		&\tabincell{c}{Fear of Spider\\and Cockroach}&Maja Wrzesien et al. \cite{RE-36}&Therapeutic Lamp&HMD\\\hline
		
		\multirow{12}{*}{Phobias of scene}
		&\multirow{3}{*}{Acrophobia}&S.Shahana et al.\cite{RE-9}&Live Beyond Fear&VR-BOX\\\cline{3-5}
		&\multirow{3}{*}&Reza Darooei et al. \cite{RE-37}&BarBam&HMD\\\cline{3-5}
		&\multirow{3}{*}&Erick Marchelino Suyanto et al. \cite{RE-38}&Acrophobia Simulator&VR-BOX\\\cline{2-5}
		&\multirow{3}{*}{PSTD}&Ehud Dayan et al. \cite{RE-39}&Argaman&HMD\\\cline{3-5}
		&\multirow{3}{*}&Jarrell Pair et al. \cite{RE-21}&ICT&HMD\\\cline{3-5}
		&\multirow{3}{*}&Albert Rizzo et al. \cite{RE-40}&BRAVEMIND VRET System&HMD\\\cline{2-5}
		&\multirow{2}{*}{Claustrophobia}&Morgan Bruce et al. \cite{RE-41}&Anti-PHOBIAS&HMD\\\cline{3-5}
		&\multirow{2}{*}&Vida Kabiri Rahani et al. \cite{RE-42}&VRET System&HMD\\\cline{2-5}
		&Dark Phobia&Vy Dang Ha Thanh et al. \cite{RE-43}&Room VR&VR-BOX\\\cline{2-5}
		&Storm Phobia&Cristina Botella et al. \cite{RE-44}&EMMA&CAVE\\\cline{2-5}
		&Driving Phobia&Jaye Wald et al. \cite{RE-24}&DriVR&HMD\\\cline{2-5}
		&Social Phobia&Dwi Hartanto et al. \cite{RE-45}&Memphis VR Dialogue System&HMD\\\hline
		
		\multirow{3}{*}{\tabincell{c}{Phobias of\\Physiological\\Phenomenon}}
		&Vertigo Phobia&\multicolumn{3}{c|}{-}\\\cline{2-5}
		&Asphyxia Phobia&\multicolumn{3}{c|}{-}\\\cline{2-5}
		&Vomit Phobia&\multicolumn{3}{c|}{-}\\\hline
			
	\end{tabular}
\end{table*}

\subsection{Serious games for phobias of object}
Phobias of object refer to the patients in the face of certain objects where they show a strong sense of fear. These objects include living objects (snakes, spiders, cockroaches, dogs, etc.), and inanimate objects (water, fire, blood, etc.). The clinical response includes the patient's involuntary avoidance behavior, sometimes accompanied by the rise in blood pressure or heart rate. For this type of phobias, serious games make patients interact with the things they fear as much as possible to suppress avoidance behavior.

Philip Lindner et al. \cite{RE-31} proposed a serious game that can be used to treat spider phobia, which uses two ways to induce fear stimulation in patients. First, the game sets different levels, level 1 is to watch the spider, level 2 is to catch the spider later, level 3 is to kill the spider. Second, the spider became more realistic from the original cartoon spider to the "real" black widow spider as the level of the game increased. The treatment showed that the game was effective in reducing patients' avoidance behavior. Spider World \cite{RE-32} has done patients' fear stimulation by letting spiders randomly fall on the patients, supplemented by Spider plush toys to increase the sense of touch. David Bel Lang et al. \cite{RE-33} proposed a serious game that can be used to treat spider phobia. To complete the interaction, the spider moves to the designated position to induce fear stimulation in patients. ARcockroach \cite{RE-34} is a serious game that can cure cockroach phobia. Patients are asked to kill the cockroaches and throw their dead bodies into the trash. Joao Petersen et al. \cite{RE-35} proposed a serious game aimed at blood phobia. In this game, the patients play the role of a detective. Their task is to find the truth of the murder scene by collecting clues and items related to blood. As the game progresses, there will be more intense blood stimulation.

In addition to these serious games that can only treat a single phobia, there is also a class of serious games that can be used to treat many different phobias. For example, Maja Wrzesien et al. \cite{RE-36} proposed a serious game, which could be used to treat both cockroach phobia and arachnophobia. These virtual animals would be placed in the hands of patients, and patients would be asked to interact with them and stare at them. 

\subsection{Serious games for phobias of scene}
Phobia of the scene is a condition in which the patient shows a strong sense of fear in the face of certain scenes. These scenes include natural environment (heights, thunderstorms, dark, etc.), and situational (flying, elevators or lifts, enclosed spaces, etc.). These serious games include acrophobia, claustrophobia, dark phobia, social phobia, and Post Traumatic Stress Disorder(PTSD). Their clinical symptoms include emotional loss, anxiety, subjective assessment of fear, and strong impulse to run away, accompanied by increased heart rate, blood pressure, and even syncope. Therefore, the serious game designed for this type of phobia should focus on improving patients' subjective cognition of the situation, controlling patients' emotions, and strengthening the interaction between patients and the virtual situation.

S. Hahana et al. \cite{RE-9}, Reza Darooei et al. \cite{RE-37}, and Erick Marchelino Suyanto et al. \cite{RE-38} proposed different serious games to treat acrophobia. In these games, the virtual scene was divided into three levels. The fear stimulus increases with the increase of the level. Patients must complete the tasks specified in each scene before they can enter the next scene. These tasks include interacting with objects at the edge of the roof, on the ground, and through a single-plank bridge with rapid water flow. The purpose is to effectively improve the patient's subjective perception of the situation and tension. Ehud Dayan et al. \cite{RE-39}, Jarrell Pair et al. \cite{RE-21}, and Albert Rizzo et al. \cite{RE-40} proposed serious games for the treatment of PTSD. These games created a virtual scene, which contains the shelling, shooting, carry the body, deal with serious injuries teammates and are unable to offer civilian help. The patient will be asked to complete these events. The results also showed that the game is effective in improving patients' uncontrolled emotions and reducing the frequency of nightmares. Morgan Bruce et al. \cite{RE-41} and Vida Kabiri Rahani et al. \cite{RE-42} proposed serious games for the treatment of claustrophobia. Patients are placed in a virtual enclosed room, and they need to explore every corner of the room. The results of the evaluation showed that the anxiety of the patients was effectively reduced in this way.

In addition to the common phobias of scene mentioned above, there are several other serious games for other phobias of scene. For example, Room VR \cite{RE-43} is a serious game for treating dark phobia. In this game, patients explore the room under the guidance of the cross line, and affect the people's emotional response is affected by changing the color of the room. Cristina Botella et al. \cite{RE-44} proposed a serious game to treat storm phobia. In this game, there are five different scenes, which are respectively related to the patients' emotions. It improved the patients' out of emotions control in the face of fear stimulation through selecting different scenes. Jaye Wald et al. \cite{RE-24} proposed a serious game used to treat driving phobia. The game consists of four scenes that the patient passes through successively by driving a car. Its purpose is to improve the anxiety of patients during driving. Dwi Hartanto et al. \cite{RE-45} proposed a serious game for treating social phobia. The game provides 19 different virtual reality social scenarios in which patients are asked to interact and talk with the virtual characters in the game. All of the social scenarios are designed to arouse the patients' social anxiety and achieve the effect of exposure therapy.

\subsection{Serious games of phobias of physiological phenomenon}
Phobias of the physiological phenomenon are the fear of some unpleasant physical reaction of the patient, such as vomiting and suffocation. Through the above two types of SGPT analysis, we can find that their treatment mechanism is through the construction of a virtual environment, and then patients are required to interact with these virtual scenes. The purpose is to do patients' fear stimulation, which to obtain the effect of virtual reality exposure therapy. However, if serious games can be used to treat phobias of physiological phenomenon, it is necessary to design virtual situations similar to physiological responses, which are difficult to achieve. As a result, there are hardly any serious games for this type of phobia.

\section{The main technologies used in serious games for phobia therapy}
\begin{figure*}[t!]
	\centering
	\includegraphics[width=15cm,height=4.5cm]{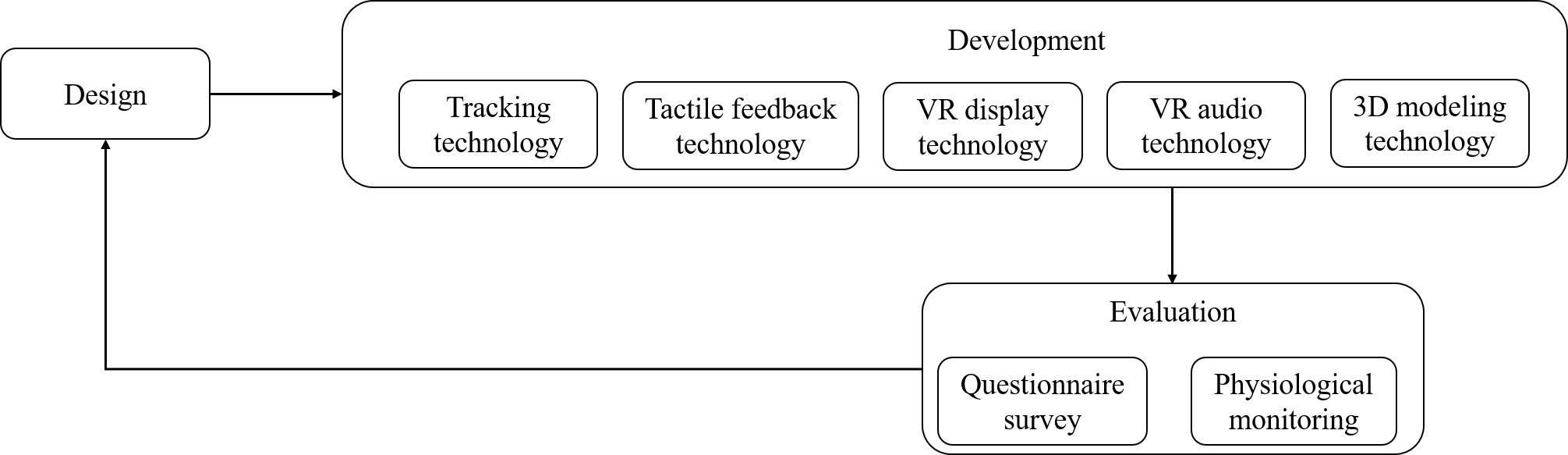}\\
	\caption{The combination of model game development process and SGPT technology }
	\label{Fig-2}
\end{figure*}

Virtual reality exposure therapy based on serious games is already widely used to treat phobias. The main mechanism of this therapy is to construct virtual fear scenarios instead of real fear scenarios, so the patient can experience the sense of “immersion”. However, most of the existing literature focuses final treatment result of SGPT. There is no provide a detailed analysis of the game development process and the technologies involved in SGPT. This review provides a detailed overview of the major technologies involved in the SGPT development process from the perspective of the Game Development Life Cycle (GDLC). In particular, our work aims to summarize the key technologies of the development process of SGPT through a new way of thinking. Its purpose to provide reference and guidance for researchers to develop SGPT. The effectiveness of this ideas and the technologies involved in each stage need further research and exploration. For clarity, Fig. \ref{Fig-2} shows a model of the combination of game development and SGPT technology.

\begin{table*}\normalsize
	\newcommand{\tabincell}[2]{\begin{tabular}{@{}#1@{}}#2\end{tabular}}
	\renewcommand\arraystretch{1.8}
	
	\centering
	\caption{Main stages of Game development process}
	\label{Tab-3}
	\begin{tabular}{|c|m{2.5cm}<{\centering}|m{2cm}<{\centering}|m{2cm}<{\centering}|m{2cm}<{\centering}|}\hline
		{\backslashbox{Stage}{GDLC name}} &{Blitz Games Studios} & {Arnold Hendrick} & {Doopler Interactive} & {Heather Chandler}\\\hline
		{Initial}&{Pitching}&{-}&{-}&{-}\\\hline
		\multirow{2}{*}{Design}&\multirow{2}{*}{Pre-production}&{Prototype}&\multirow{2}{*}{Design}&\multirow{2}{*}{Pre-production}\\\cline{3-3}
		\multirow{2}{*}{}&\multirow{2}{*}{}&{Pre-production}&\multirow{2}{*}{}&\multirow{2}{*}{}\\\hline
		\multirow{2}{*}{Development}&\multirow{2}{*}{Main production}&\multirow{2}{*}{Production}&{Develop}&\multirow
		{2}{*}{Production}\\\cline{4-4}
		\multirow{2}{*}{}&\multirow{2}{*}{}&\multirow{2}{*}{}&{Evaluate}&\multirow{2}{*}{}\\\hline
		\multirow{2}{*}{Evalution}&{Alpha}&\multirow{2}{*}{Beta}&{Test}&\multirow{2}{*}{Testing}\\\cline{2-2}\cline{4-4}
		\multirow{2}{*}{}&{Beta}&\multirow{2}{*}{}&{Review release}&\multirow{2}{*}{}\\\hline
		{Test}&{Master}&{Live}&{Release}&{Post-poduction}\\\hline				
	\end{tabular}
\end{table*}

GDLC is a guide to the game development process designed to ensure the successful development of a high-quality game. Various game developers have proposed a variety of GDLC. For example, Blitz Gams Studios \cite{RE-46} proposed a GDLC with six stages, Joshua McGrath \cite{RE-47} defined a GDLC with seven stages, Arnold Hendrick \cite{RE-48} proposed a GDLC with six stages, and Heather Chandler \cite{RE-49} proposed a GDLC with four stages. The specific phases of each GDLC are shown in TABLE \ref{Tab-3} . Comparing different GDLC, the game development process mainly includes three stages: design, development, and evaluation.

\subsection{Stage of design}
The design stage is the foundation of the whole serious game development. It is a process of creating the initial game design and game concept. We need to design a reasonable game model based on the patient's fear type and clinical manifestations. The ultimate goal is to develop a game based on this design principle that is appropriate for its target audience and allows participants to engage fully in serious games, which can be an effective tool for learning and health improvement. As the MDA framework puts it for game design: aesthetic, dynamic, and mechanical \cite{RE-65}. Based on this framework, we will illustrate the importance of design to SGPT development.

Aesthetics refers to how to make a game more enjoyable. In order to assess whether a game is fun, there are eight game aesthetic objectives within the MDA framework: Sensation, Fantasy, Narrative, Challenge, Fellowship, Discovery, Expression, and Submission. A serious game can have multiple aesthetic goals, and many different aesthetic goals can be properly combined to create the most "fun" game. Dynamic models produce aesthetic experiences. In MDA framework, there are also four typical dynamic models: challenging experience, companion experience, Expression experience, and Narrative/Dramatic experience. Mechanics are the various actions, behaviors, and control mechanisms provided to the player in the context of a game. The mechanics work together with the game content (levels, assets, etc.) to support the overall gameplay dynamics.

For example, ICT \cite{RE-21} is a serious game for the treatment of PTSD. In order to improve the immersion and realism of the game, the game uses olfactic and tactile stimuli, and there are up to six game scenarios. When participants choose different scenarios, they can either complete the task alone or together. Therefore, it is a challenge based, sensory, exploration, partner as one of the serious game. Participants need to complete specific tasks within a limited time or in a limited situation, and have challenging experiences. It includes mechanics for shooting enemies, being attacked, rescuing teammates, and burying bodies. Similarly, First Insights \cite{RE-35} is a game for the treatment of blood phobia. Aesthetically, it is a serious game featuring exploration, narrative, fantasy, and challenge. Dynamic, the participants play a detective, investigating and collecting clues at a murder scene, and eventually discovering the truth, so it has a narrative experience. Mechanical, it includes collecting clues, examining objects, reading information, and so on.

Therefore, from the point of view of game designers, when designing and developing an SGPT, we first need to identify the aesthetic goals of the game based on the needs of the target audience, then find the dynamic model behind the aesthetic goals, and finally we need to design the game mechanics and game content based on both of them. In order to provide clear guidance for the next step in game development.

\subsection{Stage of development}
\begin{figure*}[t!]
	\centering
	\includegraphics[width=16.52cm,height=8.05cm]{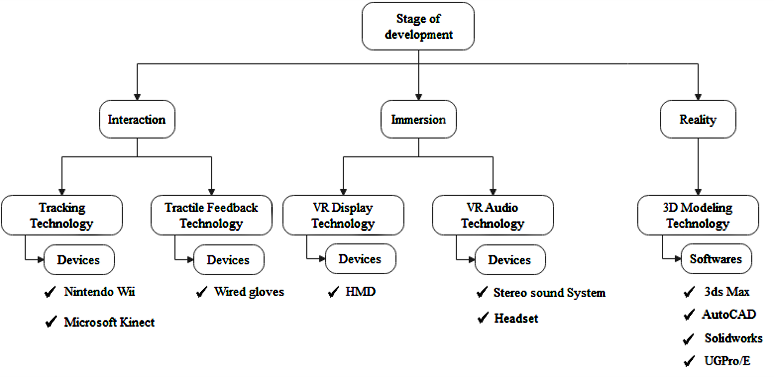}\\
	\caption{The technologies and devices involved in stage of development}
	\label{Fig-3}
\end{figure*}

The development stage is important along the process of game development. Its purpose is that to develop the SGPT according to the analysis and design in the early stage. According to the technical analysis and demand analysis in the design stage, SGPT needs to give patients the experience of interaction, realistic, and immersion. The technologies and devices involved in these three phases are shown in Fig. \ref{Fig-3}.

\subsubsection{Interaction}It refers to the user's manipulability of objects in the virtual scene and the natural degree of feedback from the environment. As the link between patients and serious games, its performance directly determines the user experience and the effectiveness of the game. This performance is mainly determined by recognition tracking technology and tactile feedback technology.

\textbf{Tracking technology}: In the virtual reality system, the used devices have to be able to read and process natural movement data to complete the tracking of patients' movements and positions. One is gesture recognition, represented by the HTC Vive and Oculus Rift motion controllers.For example, the literature \cite{RE-37} proposed a serious game to cure acrophobia, in which they have to walk across a room to an open balcony door and use the Oculus Rift motion controller to touch the balcony railings and grab hanging objects.In First Sights \cite{RE-35} In order to collect clues at the murder scene, participants used the HTC Vive motion controller to pick up broken wine glasses, grab vases, and wipe blood.The second is location tracking.Microsoft Kinect is the most famous device to use this technology. It measures the distance between the sensor and the patient using CMOS infrared sensor. Besides, it defines 48 nodes for each user and uses depth sensors to obtain the distance between each node and the sensor of the patient.These two sensors work together to realize the recognition
of patient position and movement.For example, A game for the treatment of acrophobia proposed in literature \cite{RE-38}. The game features three level environments: rivers, cities and mountains. In the final level, the player is asked to cross a cliff. Kinect captures the movement of the player to determine whether they are in a safe position or falling off the cliff. If they fall off the cliff, the player will return to the starting position until they have successfully crossed the cliff. 

\textbf{Tractile feedback technology}: In a virtual environment, people need to interact with virtual objects, and haptic feedback technology can induce fear stimulation in patients and improve the effectiveness of serious games. Tactile feedback is achieved by using wired gloves. Its basic principle is to install some vibrating contacts in the inner layer of gloves to simulate the sense of touch \cite{RE-54}. There are three main categories of the device: Electrical stimulation gloves which stimulate the skin by generating electrical pulse signals to achieve the purpose of tactile feedback; Inflatable gloves which can configure some tiny air bubbles in the gloves, then the miniature compression pump in the controller is connected with the air bubbles. After that, the compression pump is used to inflate and exhaust the air bubbles according to the needs, to obtain tactile stimulation; Vibratory gloves which integrates a vibrator into the glove to generate tactile feedback by generating vibratory stimuli.In addition, we can also add olfactory feedback technology to further enhance the game's immersion. For example, in the game ICT \cite{RE-21}  uses a scent palette of eight scent boxes that are triggered by the game's location sensor. For example, when a participant approaches a scene like smoke, a battle, a garden, etc., it triggers the device, which then releases the corresponding odor through fans installed around the room.

\subsubsection{Immersion}It is defined as having the same feeling as in the real world when the user is in the virtual world of the game. Technically, SGPT is immersive mainly through simulation in the real world of 3D visual and auditory stimuli. 

\textbf{3D vision technology}: It is achieved through a head-mounted display VR display technology. It displays two computer-generated virtual reality scene pictures from the perspective of each eye of the user. It uses the visual characteristics of human eyes to achieve a stereoscopic display effect. Besides, Field of View (FOV) refers to the visual field that human eyes can see without any movement, while the FOV of human eyes is as high as 180\degree \cite{RE-55}. Therefore, having a more accurate FOV creates a more human-like visual effect that improves immersion. For example, the HTC Vive and Oculus Rift have a FOV of 110\degree, while the more advanced Pimax 8K can achieve a FOV of 220\degree. However, according to the study on the use of head-mounted displays in the treatment of mental illness in \cite{RE-56}, based on cost considerations, the Emagine Z800 is the most commonly used head-mounted display, but its FOV is only 40\degree. In addition, the FOV of other types of head-mounted displays also have a large gap with that of human eyes. Since FOV is limited to headset display devices, additional technologies must be used to improve user immersion. 

\textbf{VR audio technology}: It can enable users to experience the sense of being surrounded by sound in a virtual environment. The sound can change as the user's gaze moves, and the user can even influence the way the sound behaves by changing the direction of their gaze. In terms of software, VR audio technology refers to construct a complete stereo sound system in the computer, such as the three-dimensional sound simulation system proposed in \cite{RE-57}. For example, 7.1 surround sound system is used in the literature \cite{RE-29}.In terms of hardware, the headset connects with the computer installed with a stereo sound system to achieve stereo surround, such as Quest 2 independent virtual reality headset launched by Oculus and AKG headset launched by Samsung.
\subsubsection{Reality}The key point of realistic technology is to construct realistic virtual scene models through 3D modeling technology. Vision is for people to obtain information about the most direct and most important ways from the real world. Therefore, the effects of models in games directly affect the reality of the virtual environment. Currently, virtual reality modeling technology mainly belongs to the following three ways.

\textbf{Geometry-Based Modeling and Rendering (GBMR)}: The principle of this technique is to build geometric models through precise parameters, making the whole virtual scene more real and fine. This allows users to interact with the virtual model in a three-dimensional scene. But it requires powerful computer hardware and software. 

\textbf{Image-Based Modeling and Rendering (IBMR)}: It can generate a high-quality virtual scene that reflects the real scene, without the need to complex modeling and professional graphics processing equipment. Therefore, it can quickly and easily construct a realistic virtual scene. However, it needs to obtain the image of the real scene in advance. Because the virtual objects in the scene are rendered and constructed based on the objects in the image, they are not strictly three-dimensional objects, so the user can hardly realize the interaction with the virtual scene. 

\textbf{Mixture of Geometry and Image-Based Modeling and Rendering}: This technology combines the advantages of the aforementioned two technologies. The overall visual effect of the game is completed by IBMR, and the part of the game that needs to interact with the user is realized by GBMR. Therefore, it can not only guarantee the simplicity and fidelity of model establishment but also guarantee good interactivity. 

The commonly used 3D modeling software mainly includes 3ds Max, AutoCAD, Solidworks, and UGPro/E. The more common approach is to build a certain ratio of the model through UGPro/E, SolidWorks, and other software, and then through 3ds Max and other softwares for the model to add the appropriate material, texture, lighting, and so on, which improves the fidelity of the objects.

\subsection{Stage of evalution}
The evaluation phase is the final stage of the entire game development process, It requires the patient's physical state to be evaluated after the patient plays SGPT to determine the effect of SGPT in the treatment of phobia. Traditional game evaluations are done by a combination of internal team members and third-party testers. According to the characteristics of SGPT, it can be evaluated from both subjective (questionnaire survey) and objective (physiological monitoring) aspects.
\subsubsection{Questionnaire survey}A questionnaire survey is the most common evaluation measure at present, which includes a series of evaluation measures. For example, in Subjective Units of Discomfort Scale (SUDS) \cite{RE-58}, patients are asked to assess their anxiety level when facing fear through using an 11-point scale between 0 (no anxiety) and 10 (extreme anxiety). Maladjustment Scales(MS) \cite{RE-59} asked patients to also use an 11-point scale to assess the impact of phobias on their daily lives (0: None; 10: Extreme). Behavioral Avoidance Test (BAT) \cite{RE-60} is used to evaluate the degree of avoidance of patients in the face of fear. It uses an 11-point scale to evaluate the subjective fear and avoidance degree of patients. The higher the score, the more severe the patient's condition. Therefore, for the patients with low scores, this indicates that the developed serious games have a significant effect on the treatment of phobia. On the contrary, for the patients with high scores, the developed serious games have a poor effect on the treatment of phobia. However, a questionnaire survey can be influenced by patients' subjective consciousness, so monitoring patients' physiological response is a good supplement.
\subsubsection{Physiological monitoring}Physiological monitoring is the process of measuring physiological functions through sensors and manipulating the collected data. It has been widely applied to the condition assessment of patients with phobias \cite{RE-61}. It allows a more objective way to measure the response of patients in the face of fear. Some changes in physical function can be directly associated with changes in the mood \cite{RE-62}. Studies have shown that people with phobias have abnormal brain activity, particularly in areas where fear is perceived and amplified. Therefore, Commercial electroencephalography (EEG) solutions, such as the Neurosky Mindwave, can monitor the emotional state of the patient. Such as relaxation, depression, nervousness, or others \cite{RE-63}. Besides, Galvanic Skin Response (GSR) can be used to measure the patient's excitement state, which is usually measured by two electrodes attached to on the patient's hand. These electrodes measure the current difference caused by the increase of exocrine sweat gland activity, which is usually the result of excitement \cite{RE-64}. In this way, a questionnaire survey and physiological monitoring were used together as evaluation criteria to improve the quality of SGPT.

In summary, we used three phases of GDLC in this chapter and describe the main technologies applied in different phases. There is a cyclical relationship between the stages. In the design phase, the appropriate SGPT framework is constructed according to the patient status and game engine. In the development stage, SGPT with specific functions was developed based on the requirements of mutual sense, immersion sense, and real sense. SGPT deficiencies were documented through patient feedback during the evaluation phase and then moved on to the next development process. The quality of the game can be improved by constantly iterating on the SGPT to achieve the best therapeutic effect. In addition, compared with other relevant literature, our research focuses on the technical level, which is also a significant advantage of our model.

\section{Discussion and prospect}
Virtual reality exposure therapy based on serious games has shown great important role in the treatment of phobias and many related studies have shown the effectiveness of this approach. Many serious games are used to treat phobias. Although the research on this method has been going on for a long time, VR is still an emerging technology. Game design and treatment purposes cannot achieve perfect integration. In addition, although there are already a wealth of serious games, it is a single category that does not cover a large number of different types of phobias. Given these challenges, the following directions deserve further research and development:

\begin{itemize}	
	\item The duration of treatment is too long, making patient's fatigue will inevitably increases. The patient should be able to freely choose the duration of treatment, and game should have a memory function and can to remember the patient's last play record.	
	\item For the SGPT, its type of play is usually first-person and patients are required bulky HMD. As a result, patients can feel dizzy when they play for too long, which increases the patient's physical burden due to the heavy equipment. In the future, we can develop more portable virtual display devices can be developed or develop more serious games based on CAVE can be created.
	\item Serious games for phobia therapy are expensive and poorly understood. In the face of this situation, on the one hand, it is necessary to increase popularization so that more patients understand this new type of treatment. On the other hand, the government can issue relevant subsidy policies to help hospitals be equipped with relevant equipment. This may not only increase users' awareness of the SGPT, but also enable patients to enjoy abundant medical resources to reduce their burden.
	\item Serious game development engines for phobias use game engines designed specifically for video games, and these game engines lack the technology for therapy. Therefore, a set of the game engine for SGPT can be developed, which can not only improve the richness of SGPT but also achieve the best balance between game design and therapy purpose.
	\item In the current SGPT, the physician will always pay attention to the patient's state in another treatment room while the patient is being treated. That helps to adjust the fear stimulation of the game. Therefore, physiological monitoring can be introduced into serious games. The physiological response of patients can be used as a feedback factor, so that serious games can automatically regulate fear stimulation without the participation of doctors.
	\item At present, most SGPT treatments are for patients with a single symptom.  Although there are a few of them that can be used to treat multiple patients with similar symptoms, there is no SGPT that can be used to treat patients with all symptoms. SGPT needs to be developed separately for each symptom, which increases the cost of treatment and makes the quality of the game is uneven. In the future, a system could be proposed that can be used to treat all types of phobias. Patients can choose the appropriate SGPT and game mechanics to achieve the best therapeutic effect according to their situation.	
\end{itemize}

\section{Conclusion}
SGPT has been well studied as an effective therapy for phobias. However, despite long-term research has been conducted, there is still a lack of a systematic review of the development history of SGPT. In order to fill this gap, this paper described and summarized the development history of SGPT from the perspective of VR display devices and divided it into three aspects: VR-Box, HMD, and CAVE. These studies can help relevant researchers to some extent. In addition, we found that although there are many serious games for phobia-related therapy, there is no consistent classification or generalization of these serious games in the literature. Therefore, the existing serious games has been classified into three categories according to the clinical symptoms of phobias. In the end, we divided SGPT development process into three stages based on the commercial development process of video games and the characteristics of SGPT. The main technology for each stage is described in detail. That will help researchers effectively develop SGPT and achieve the optimal balance between game design and therapeutic purpose. In conclusion, our work aims to provide a comprehensive summary of existing SGPT and provide theoretical guidance for the design of a more effective SGPT. 



\section*{Acknowledgment}
The work was supported by teaching reform of University of Science and Technology Beijing under JG2020Z08.
\bibliographystyle{unsrt}
\bibliography{Reference}

\begin{thebibliography}{10}

\bibitem{RE-1}
SE~Cassin, JH~Riskind, and NA~Rector.
\newblock Phobias.
\newblock 2012.

\bibitem{RE-2}
Ella~L Oar, Lara~J Farrell, and Thomas~H Ollendick.
\newblock Specific phobia.
\newblock In {\em Pediatric Anxiety Disorders}, pages 127--150. Elsevier, 2019.

\bibitem{RE-3}
Yujuan Choy, Abby~J Fyer, and Josh~D Lipsitz.
\newblock Treatment of specific phobia in adults.
\newblock {\em Clinical psychology review}, 27(3):266--286, 2007.

\bibitem{RE-4}
Dennis Ougrin.
\newblock Efficacy of exposure versus cognitive therapy in anxiety disorders:
  systematic review and meta-analysis.
\newblock {\em BMC psychiatry}, 11(1):1--13, 2011.

\bibitem{RE-5}
Thomas~H Ollendick and Thompson~E Davis~III.
\newblock One-session treatment for specific phobias: a review of {\"o}st's
  single-session exposure with children and adolescents.
\newblock {\em Cognitive behaviour therapy}, 42(4):275--283, 2013.

\bibitem{RE-6}
Alexander Miloff, Philip Lindner, William Hamilton, Lena Reuterski{\"o}ld,
  Gerhard Andersson, and Per Carlbring.
\newblock Single-session gamified virtual reality exposure therapy for spider
  phobia vs. traditional exposure therapy: study protocol for a randomized
  controlled non-inferiority trial.
\newblock {\em Trials}, 17(1):60, 2016.

\bibitem{RE-7}
Theresa~M Fleming, Lynda Bavin, Karolina Stasiak, Eve Hermansson-Webb, Sally~N
  Merry, Colleen Cheek, Mathijs Lucassen, Ho~Ming Lau, Britta Pollmuller, and
  Sarah Hetrick.
\newblock Serious games and gamification for mental health: current status and
  promising directions.
\newblock {\em Frontiers in psychiatry}, 7:215, 2017.

\bibitem{RE-8}
Azucena Garcia-Palacios, Hunter~G Hoffman, Sheree Kwong~See, AMY Tsai, and
  Cristina Botella.
\newblock Redefining therapeutic success with virtual reality exposure therapy.
\newblock {\em CyberPsychology \& Behavior}, 4(3):341--348, 2001.

\bibitem{RE-9}
S~Shahana Sharmili and R~Kanagaraj.
\newblock Live beyond fear: A virtual reality serious game platform to overcome
  phobias.
\newblock In {\em 2020 5th International Conference on Devices, Circuits and
  Systems (ICDCS)}, pages 336--339. IEEE, 2020.

\bibitem{RE-10}
Mario Gutierrez, Fr{\'e}d{\'e}ric Vexo, and Daniel Thalmann.
\newblock {\em Stepping into virtual reality}.
\newblock Springer Science \& Business Media, 2008.

\bibitem{RE-11}
Jo{\~a}o~P Costa, James Robb, and Lennart~E Nacke.
\newblock Physiological acrophobia evaluation through in vivo exposure in a vr
  cave.
\newblock In {\em 2014 IEEE Games Media Entertainment}, pages 1--4. IEEE, 2014.

\bibitem{RE-12}
Alfons~O Hamm.
\newblock Phobias across the lifespan.
\newblock 2015.

\bibitem{RE-43}
Vy~Dang~Ha Thanh, Ondris Pui, and Martin Constable.
\newblock Room vr: a vr therapy game for children who fear the dark.
\newblock In {\em SIGGRAPH Asia 2017 Posters}, pages 1--2. 2017.

\bibitem{RE-20}
NC~Maatjes.
\newblock The treatment of phobias using virtual reality.
\newblock In {\em 3rd Twente Student Conference on IT, Enschede, June}.
  Citeseer, 2005.

\bibitem{RE-21}
Jarrell Pair, Brian Allen, Matthieu Dautricourt, Anton Treskunov, Matt Liewer,
  Ken Graap, and Greg Reger.
\newblock A virtual reality exposure therapy application for iraq war post
  traumatic stress disorder.
\newblock In {\em IEEE Virtual Reality Conference (VR 2006)}, pages 67--72.
  IEEE, 2006.

\bibitem{RE-22}
Rosa~Mar{\'\i}a Ba{\~n}os, Cristina Botella, Concepci{\'o}n Perpi{\~n}{\'a},
  Mariano Alca{\~n}iz, Jose~Antonio Lozano, Jorge Osma, and Myriam Gallardo.
\newblock Virtual reality treatment of flying phobia.
\newblock {\em IEEE Transactions on Information Technology in Biomedicine},
  6(3):206--212, 2002.

\bibitem{RE-23}
St{\'e}phane Bouchard, Sophie C{\^o}t{\'e}, Julie St-Jacques, Genevi{\`e}ve
  Robillard, and Patrice Renaud.
\newblock Effectiveness of virtual reality exposure in the treatment of
  arachnophobia using 3d games.
\newblock {\em Technology and health care}, 14(1):19--27, 2006.

\bibitem{RE-24}
Jaye Wald and Steven Taylor.
\newblock Efficacy of virtual reality exposure therapy to treat driving phobia:
  a case report.
\newblock {\em Journal of behavior therapy and experimental psychiatry},
  31(3-4):249--257, 2000.

\bibitem{RE-25}
Sherazade Shunnaq and Mateus Raeder.
\newblock Virtualphobia: A model for virtual therapy of phobias.
\newblock In {\em 2016 XVIII Symposium on Virtual and Augmented Reality (SVR)},
  pages 59--63. IEEE, 2016.

\bibitem{RE-26}
Jessica~S Ortiz, Paola~M Velasco, Washington~X Quevedo, Marcelo {\'A}lvarez,
  Jorge~S S{\'a}nchez, Christian~P Carvajal, Luis~F Cepeda, and V{\'\i}ctor~H
  Andaluz.
\newblock Realism in audiovisual stimuli for phobias treatments through virtual
  environments.
\newblock In {\em International Conference on Augmented Reality, Virtual
  Reality and Computer Graphics}, pages 188--201. Springer, 2017.

\bibitem{RE-27}
Miroslav Musalek and Lubomir Vasek.
\newblock Possibilities of using virtual reality as a means for therapy from
  fear of spiders.
\newblock In {\em MATEC Web of Conferences}, volume 292, page 01041. EDP
  Sciences, 2019.

\bibitem{RE-28}
Remi~Jounghuem Kwon.
\newblock {\em Anxiety activating virtual environments for investigating social
  phobias}.
\newblock PhD thesis, University of Warwick, 2010.

\bibitem{RE-29}
Daniel Gromer, Oct{\'a}via Madeira, Philipp Gast, Markus Nehfischer, Michael
  Jost, Mathias M{\"u}ller, Andreas M{\"u}hlberger, and Paul Pauli.
\newblock Height simulation in a virtual reality cave system: validity of fear
  responses and effects of an immersion manipulation.
\newblock {\em Frontiers in human neuroscience}, 12:372, 2018.

\bibitem{RE-30}
Merel Krijn, Paul~MG Emmelkamp, Roeline Biemond, Claudius de~Wilde de~Ligny,
  Martijn~J Schuemie, and Charles~APG van~der Mast.
\newblock Treatment of acrophobia in virtual reality: The role of immersion and
  presence.
\newblock {\em Behaviour research and therapy}, 42(2):229--239, 2004.

\bibitem{RE-19}
Maria Abdullah and Zubair~Ahmed Shaikh.
\newblock An effective virtual reality based remedy for acrophobia.
\newblock {\em International Journal of Advanced Computer Science and
  Applications.—2018.—9 (6)}, 2018.

\bibitem{RE-31}
Philip Lindner, Alexander Rozental, Alice Jurell, Lena Reuterski{\"o}ld,
  Gerhard Andersson, William Hamilton, Alexander Miloff, and Per Carlbring.
\newblock Experiences of gamified and automated virtual reality exposure
  therapy for spider phobia: Qualitative study.
\newblock {\em JMIR Serious Games}, 8(2):e17807, 2020.

\bibitem{RE-32}
Merel Krijn, Paul~MG Emmelkamp, Roeline Biemond, Claudius de~Wilde de~Ligny,
  Martijn~J Schuemie, and Charles~APG van~der Mast.
\newblock Treatment of acrophobia in virtual reality: The role of immersion and
  presence.
\newblock {\em Behaviour research and therapy}, 42(2):229--239, 2004.

\bibitem{RE-33}
D~Bel~Lang.
\newblock Augmented reality phobia treatment including biofeedback.
\newblock 2018.

\bibitem{RE-34}
MC~Juan, Cristina Botella, M~Alcaniz, R~Banos, C~Carrion, M~Melero, and
  Jos{\'e}~Antonio Lozano.
\newblock An augmented reality system for treating psychological disorders:
  application to phobia to cockroaches.
\newblock In {\em Third IEEE and ACM International Symposium on Mixed and
  Augmented Reality}, pages 256--257. IEEE, 2004.

\bibitem{RE-35}
Jo{\~a}o Petersen, V{\'\i}tor Carvalho, Jo{\~a}o~Tiago Oliveira, and Eva
  Oliveira.
\newblock A serious game for hemophobia treatment phobos: First insights.
\newblock In {\em Interactivity, Game Creation, Design, Learning, and
  Innovation}, pages 231--236. Springer, 2018.

\bibitem{RE-36}
Maja Wrzesien, Mariano Alca{\~n}iz, Cristina Botella, Jean-Marie Burkhardt,
  Juana Bret{\'o}n-L{\'o}pez, Mario Ortega, and Daniel~Beneito Brotons.
\newblock The therapeutic lamp: treating small-animal phobias.
\newblock {\em IEEE computer graphics and applications}, 33(1):80--86, 2013.

\bibitem{RE-37}
Reza Darooei, Alireza Vard, and Hossein Rabbani.
\newblock Barbam: a new arcophobia virtual reality game.
\newblock In {\em 2019 International Serious Games Symposium (ISGS)}, pages
  48--53. IEEE, 2019.

\bibitem{RE-38}
Erick~Marchelino Suyanto, Denny Angkasa, Harfondy Turaga, and Rhio Sutoyo.
\newblock Overcome acrophobia with the help of virtual reality and kinect
  technology.
\newblock {\em Procedia computer science}, 116:476--483, 2017.

\bibitem{RE-39}
Ehud Dayan.
\newblock Argaman: Rapid deployment virtual reality system for ptsd
  rehabilitation.
\newblock In {\em 2006 International Conference on Information Technology:
  Research and Education}, pages 34--38. IEEE, 2006.

\bibitem{RE-40}
Albert Rizzo, Michael~J Roy, Arno Hartholt, Michelle Costanzo, Krista~Beth
  Highland, Tanja Jovanovic, Seth~D Norrholm, Chris Reist, Barbara Rothbaum,
  and JoAnn Difede.
\newblock Virtual reality applications for the assessment and treatment of
  ptsd.
\newblock In {\em Handbook of Military Psychology}, pages 453--471. Springer,
  2017.

\bibitem{RE-41}
Iulia-Cristina St{\u{a}}nic{\u{a}}, Maria-Iuliana Dasc{\u{a}}lu, Alin
  Moldoveanu, and Florica Moldoveanu.
\newblock An innovative solution based on virtual reality to treat phobia.
\newblock {\em Int J Interact Worlds}, 2017:1--13, 2017.

\bibitem{RE-42}
Morgan Bruce and Holger Regenbrecht.
\newblock A virtual reality claustrophobia therapy system-implementation and
  test.
\newblock In {\em 2009 IEEE Virtual Reality Conference}, pages 179--182. IEEE,
  2009.

\bibitem{RE-44}
Cristina Botella, Rosa~Mar{\'\i}a Ba{\~n}os, Bel{\'e}n Guerrero, Azucena
  Garc{\'\i}a-Palacios, Soledad Quero, and Mariano~Alca{\~n}iz Raya.
\newblock Using a flexible virtual environment for treating a storm phobia.
\newblock {\em PsychNology Journal}, 4(2):129--144, 2006.

\bibitem{RE-45}
Dwi Hartanto, Willem-Paul Brinkman, Isabel~L Kampmann, Nexhmedin Morina,
  Paul~GM Emmelkamp, and Mark~A Neerincx.
\newblock Home-based virtual reality exposure therapy with virtual health agent
  support.
\newblock In {\em International Symposium on Pervasive Computing Paradigms for
  Mental Health}, pages 85--98. Springer, 2015.

\bibitem{RE-46}
Saiqa Aleem, Luiz~Fernando Capretz, and Faheem Ahmed.
\newblock Game development software engineering process life cycle: a
  systematic review.
\newblock {\em Journal of Software Engineering Research and Development},
  4(1):6, 2016.

\bibitem{RE-47}
J~McGrath.
\newblock The game development lifecycle: A theory for the extension of the
  agile project methodology.
\newblock {\em Available via online. http://blog. dopplerinteractive.
  com/post/112172271166/the-gamedevelopment-lifecycle-a-theory-for-the[Accessed
  05/06/17]}, 2014.

\bibitem{RE-48}
A~Hendrick.
\newblock Project management for game development.
\newblock {\em Available via online. https://mmotidbits.
  com/2009/06/15/project-management-forgame-development/[Accessed 05/06/17]},
  2009.

\bibitem{RE-49}
Heather~Maxwell Chandler.
\newblock {\em The game production handbook}.
\newblock Jones \& Bartlett Publishers, 2009.

\bibitem{RE-65}
Robin Hunicke, Marc LeBlanc, and Robert Zubek.
\newblock Mda: A formal approach to game design and game research.
\newblock In {\em Proceedings of the AAAI Workshop on Challenges in Game AI},
  volume~4, page 1722. San Jose, CA, 2004.

\bibitem{RE-54}
David~J Sturman and David Zeltzer.
\newblock A survey of glove-based input.
\newblock {\em IEEE Computer graphics and Applications}, 14(1):30--39, 1994.

\bibitem{RE-55}
John~Marco Oscillada.
\newblock Comparison chart of fov (field of view) of vr headsets.
\newblock {\em Virtual Reality Times}, 2015.

\bibitem{RE-56}
Shaun~W Jerdan, Mark Grindle, Hugo~C van Woerden, and Maged N~Kamel Boulos.
\newblock Head-mounted virtual reality and mental health: critical review of
  current research.
\newblock {\em JMIR serious games}, 6(3):e14, 2018.

\bibitem{RE-57}
Kai-Uwe Doerr, Holger Rademacher, Silke Huesgen, and Wolfgang Kubbat.
\newblock Evaluation of a low-cost 3d sound system for immersive virtual
  reality training systems.
\newblock {\em IEEE Transactions on Visualization and Computer Graphics},
  13(2):204--212, 2007.

\bibitem{RE-58}
Joseph Wolpe.
\newblock {\em The practice of behavior therapy}.
\newblock Pergamon Press, 1990.

\bibitem{RE-59}
Enrique~Echebur{\'u}a Odriozola and Paz de~Corral.
\newblock {\em La agorafobia: Nuevas perspectivas de evaluaci{\'o}n y
  tratamiento}.
\newblock Promolibro, 1992.

\bibitem{RE-60}
Lars-G{\"o}ran {\"O}st, Paul~M Salkovskis, and Kerstin Hellstr{\"o}m.
\newblock One-session therapist-directed exposure vs. self-exposure in the
  treatment of spider phobia.
\newblock {\em Behavior Therapy}, 22(3):407--422, 1991.

\bibitem{RE-61}
John~L Reeves and Wallace~L Mealiea.
\newblock Biofeedback-assisted cue-controlled relaxation for the treatment of
  flight phobias.
\newblock {\em Journal of Behavior Therapy and Experimental Psychiatry},
  6(2):105--109, 1975.

\bibitem{RE-62}
Arthur~SP Jansen, Xay Van~Nguyen, Vladimir Karpitskiy, Thomas~C Mettenleiter,
  and Arthur~D Loewy.
\newblock Central command neurons of the sympathetic nervous system: basis of
  the fight-or-flight response.
\newblock {\em Science}, 270(5236):644--646, 1995.

\bibitem{RE-63}
Lennart~E Nacke.
\newblock Wiimote vs. controller: electroencephalographic measurement of
  affective gameplay interaction.
\newblock In {\em Proceedings of the international academic conference on the
  future of game design and technology}, pages 159--166, 2010.

\bibitem{RE-64}
Lennart~E Nacke.
\newblock An introduction to physiological player metrics for evaluating games.
\newblock In {\em Game Analytics}, pages 585--619. Springer, 2013.

\end{thebibliography}

%

\begin{IEEEbiography}[{\includegraphics[width=1in,height=1.25in,clip,keepaspectratio]{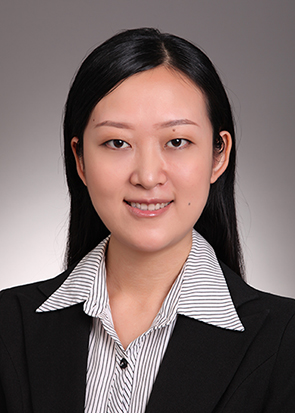}}]{Sha Li}
	was admitted to information science and technology college of hainan university in 2004 and studied communication engineering specialty. She confirmed into the information science and technology college of hainan university master of professional communication and information system in 2008 and was admitted to Beijing university of posts and telecommunications institute of information photonics and optical communications at optical engineering PhD student in 2011. During this period, he was funded by Dublin Institute of Technology International Scholarship from the Irish government. He studied in the Light Research Center of Dublin Institute of Technology. In April 2013, he was awarded by Irish President Mack. Higgins personally.She is currently an associate professor in the Department of Internet of Things and Electronic Engineering, School of Computer and Communication Engineering, University of Science and Technology Beijing. Her research interests include atmospheric optical communication, high speed analog-to-digital conversion technology, and high speed optical sampling based on time domain broadening
\end{IEEEbiography}

\begin{IEEEbiography}[{\includegraphics[width=1in,height=1.25in,clip,keepaspectratio]{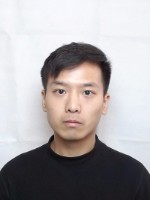}}]{Peichen Yang}
	received the B.S. degree from Henan University of Technology and currently working toward the M.S. degree in the School of Computer and Communication Engineering,University of Science and Technology Beijing,China. His current research focuses the application of serious games.
\end{IEEEbiography}

\begin{IEEEbiography}[{\includegraphics[width=1in,height=1.25in,clip,keepaspectratio]{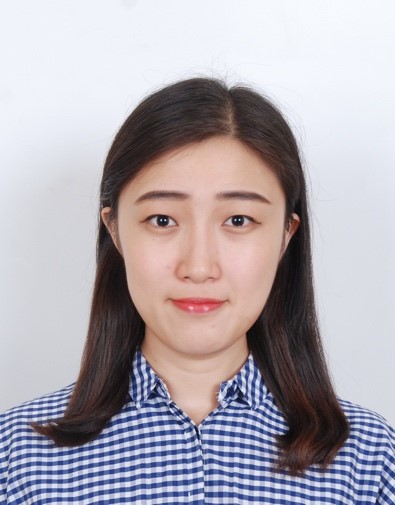}}]{Rongyang Li}
	received her B.S. degree from Hebei Normal University of Science and Technology in 2018 and her M.S. degree from University of Science and Technology Beijing in January 2021. She will begin to pursue her PhD. Degree in September 2021 at the School of Computer and Communication Engineering, University of Science and Technology Beijing, China. Her current research focuses the user experience of games.
\end{IEEEbiography}

\begin{IEEEbiography}[{\includegraphics[width=1in,height=1.25in,clip,keepaspectratio]{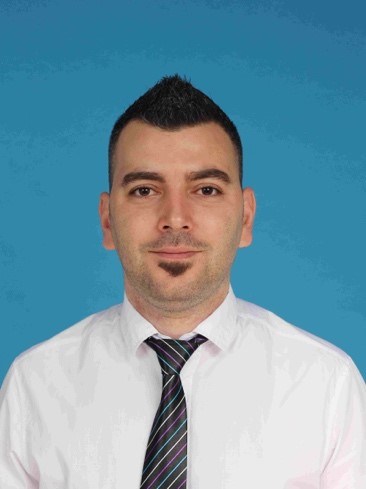}}]{Fadi Farha}
	received his MS degree and currently working toward Ph.D. degree in the School of Computer and Communication Engineering, University of Science and Technology Beijing, China. His current research interests include Physical Unclonable Function (PUF), Smart Home, Security Solutions, ZigBee, Computer Architecture and Hardware Security.
\end{IEEEbiography}

\begin{IEEEbiography}[{\includegraphics[width=1in,height=1.25in,clip,keepaspectratio]{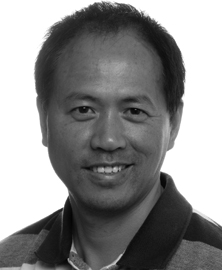}}]{Jianguo Ding}
holds the degree of a Doctorate in Engineering (Dr.-Ing.) from the faculty of mathematics and computer Science in University of Hagen, Germany. He is currently a senior lecturer at school of informatics in University of Skovde, Sweden. His current research interests include distributed systems management and control, intelligent technology, probabilistic reasoning and critical infrastructure protection.
\end{IEEEbiography}

\begin{IEEEbiography}[{\includegraphics[width=1in,height=1.25in,clip,keepaspectratio]{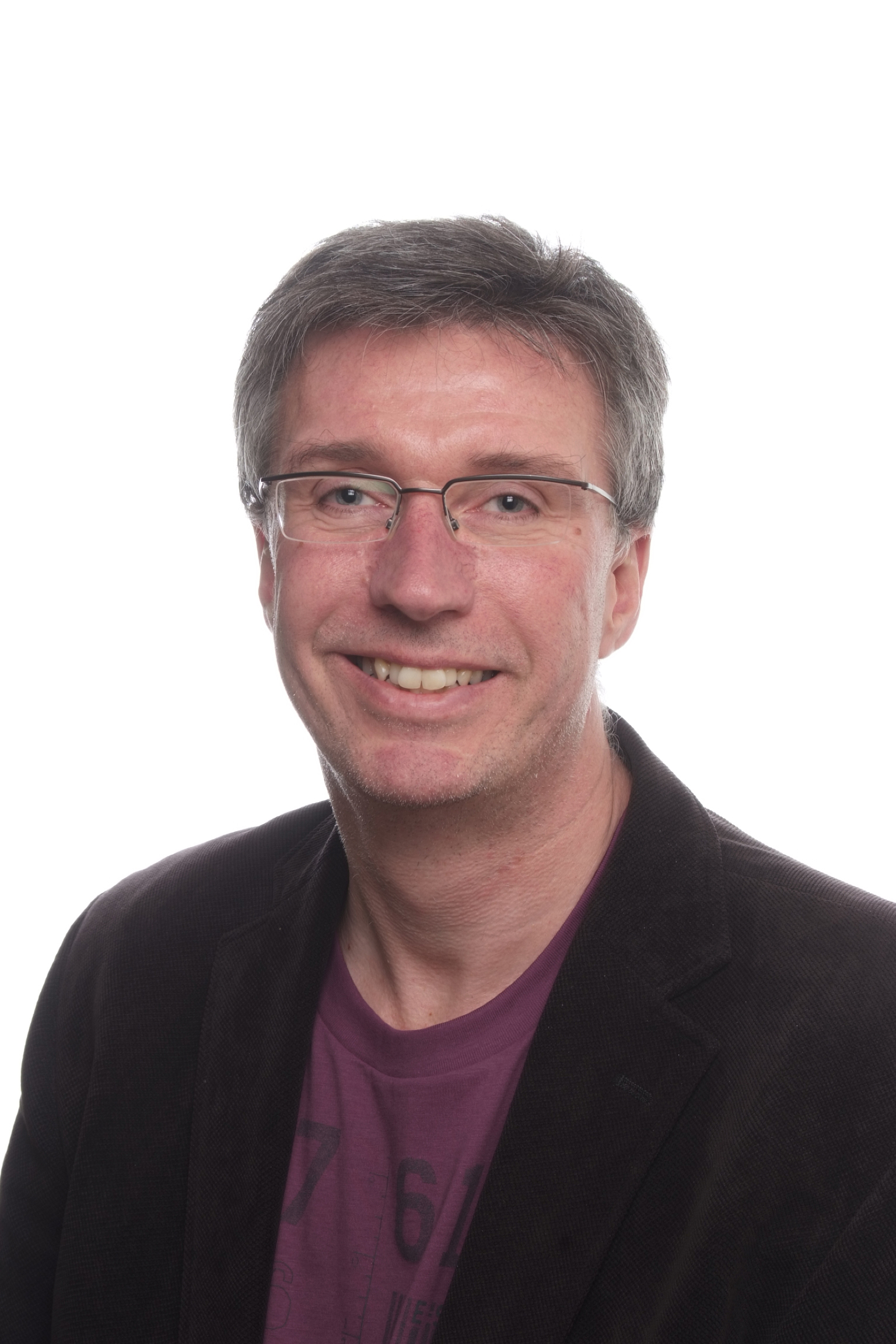}}]{Per Backlund}
	holds a PhD from Stockholm University and is currently a Professor of Information Technology at University of Skovde. He has been an active researcher in the field of serious games since 2005 with a specialization in game based and simulation based training. Professor Backlund has had the role of project manager and principal investigator in several research projects in serious games applications for different application areas such as traffic education, rescue services training and prehospital medicine.
\end{IEEEbiography}

\begin{IEEEbiography}[{\includegraphics[width=1in,height=1.25in,clip,keepaspectratio]{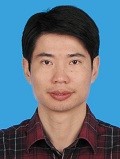}}]{Huansheng Ning}
received his B.S. degree from Anhui University in 1996 and his Ph.D. degree from Beihang University in 2001. He is currently a Professor and Vice Dean with the School of Computer and Communication Engineering, University of Science and Technology Beijing and China and Beijing Engineering Research Center for Cyberspace Data Analysis and Applications, China, and the founder and principal at Cybermatics and Cyberspace International Science and Technology Cooperation Base. He has authored several books and over 70 papers in journals and at international conferences/workshops. He has been the Associate Editor of IEEE Systems Journal and IEEE Internet of Things Journal, Chairman (2012) and Executive Chairman (2013) of the program committee at the IEEE international Internet of Things conference, and the Co-Executive Chairman of the 2013 International cyber technology conference and the 2015 Smart World Congress. His awards include the IEEE Computer Society Meritorious Service Award and the IEEE Computer Society Golden Core Member Award. His current research interests include Internet of Things, Cyber Physical Social Systems, electromagnetic sensing and computing.
\end{IEEEbiography}

\end{document}